\documentclass{article}

\evensidemargin=\oddsidemargin
\def\Title#1#2#3{%
    \baselineskip=18pt
    \begin{center}
          {\large\bf{#1} \\ }
          \bigskip\bigskip
          {#2} \\
          {#3} \\
    \end{center}}
\long\def\Abstract#1{%
         \bigskip
         \parbox{0.93\textwidth}{%
                 \begin{center}
                       {\bf Abstract} \\
                 \end{center}
                 \medskip{\baselineskip=14pt #1}
                 \vss}
         \bigskip}

\makeatletter
\renewcommand{\section}%
 {\@startsection{section}{1}{0pt}%
  {-3.25ex plus -1ex minus -.2ex}{1.5ex plus .2ex}%
  {\vspace*{5mm}\raggedright\large\bf }}
\renewcommand{\subsection}%
 {\@startsection{subsection}{2}{0pt}%
  {-2.25ex plus -.5ex minus -.2ex}{-1.5ex plus -.2ex}{\bf }}
\renewcommand{\subsubsection}%
 {\@startsection{subsubsection}{3}{0pt}%
  {-1.25ex plus -.2ex minus -.1ex}{-1.2ex plus -.2ex}{\bf }}

\@addtoreset{equation}{section}

\makeatother

\begin{document}

\Title{The role of BRST charge as a generator of gauge transformations\\
in quantization of gauge theories and Gravity}%
{T. P. Shestakova}%
{Department of Theoretical and Computational Physics,
Southern Federal University,\\
Sorge St. 5, Rostov-on-Don 344090, Russia \\
E-mail: {\tt shestakova@sfedu.ru}}

\Abstract{In the Batalin -- Fradkin -- Vilkovisky approach to quantization of gauge theories a principal role is given to the BRST charge which can be constructed as a series in Grassmannian (ghost) variables with coefficients given by generalized structure functions of constraints algebra. Alternatively, the BRST charge can be derived making use of the Noether theorem and global BRST invariance of the effective action. In the case of Yang -- Mills fields the both methods lead to the same expression for the BRST charge, but it is not valid in the case of General Relativity. It is illustrated by examples of an isotropic cosmological model as well as by spherically-symmetric gravitational model which imitates the full theory of gravity much better. The consideration is based on Hamiltonian formulation of General Relativity in extended phase space. At the quantum level the structure of the BRST charge is of great importance since BRST invariant quantum states are believed to be physical states. Thus, the definition of the BRST charge at the classical level is inseparably related to our attempts to find a true way to quantize Gravity.}

\section{Introduction}
In the Batalin -- Fradkin -- Vilkovisky (BFV) approach to quantization of gauge theories \cite{BFV1,BFV2,BFV3} a principal role is given to the BRST charge since BRST invariant quantum states are believed to be physical states. As I shall demonstrate, in the case of gravity one meets the problem how the BRST charge should be defined and, therefore, what are physical states. The aim of my talk is to attract attention to this problem.

Let me start from well-known things. In the BFV approach the BRST charge can be constructed as a series in Grassmannian (ghost) variables with coefficients given by generalized structure functions of constraints algebra \cite{Hennaux}:
\begin{equation}
\label{BFV BRST}
\Omega_{BFV}=\int d^3x\left(c^{\alpha}U^{(0)}_{\alpha}
 +c^{\beta}c^{\gamma}U^{(1)\alpha}_{\gamma\beta}\bar\rho_{\alpha}+\ldots\right)
\end{equation}
$c^{\alpha}$, $\bar\rho_{\alpha}$ are the BFV ghosts and their conjugate momenta, $U^{(n)}$ are $n$th order structure functions, while zero order structure functions $U^{(0)}_{\alpha}=G_{\alpha}$ are Dirac secondary constraints. In quantum theory physical states are those annihilated by the BRST charge $\hat\Omega$:
\begin{equation}
\label{quant BRST}
\hat\Omega|\Psi\rangle=0.
\end{equation}
It can be proved that the condition (\ref{quant BRST}) is equivalent to the quantum version of constraints:
\begin{equation}
\label{quant Dirac}
\hat G_{\alpha}|\Psi\rangle=0.
\end{equation}
The proof \cite{Hennaux} is essentially based upon the statement that any set of constraints is equivalent (at the classical level) to another set of strongly commuting constraints. Then the expansion (\ref{BFV BRST}) is reduced to the first term only. The proof is formal and ignores such problems as operator ordering. However, we shall not discuss its details here.

Let us note that there exist another way to construct the BRST charge making use of global BRST symmetry and the Noether theorem. In the case of Yang -- Mills fields this method leads to the same expression for the BRST charge as the BFV prescription (\ref{BFV BRST}). For example, let us consider the Faddeev -- Popov action for the Yang -- Mills fields in the Lorentz gauge
\begin{equation}
\label{YMF}
S_{YM}=\int d^4x\left[-\frac14F_{\mu\nu}^aF_a^{\mu\nu}
   -i\bar{\theta}_a\partial^{\mu}D_{\mu}\theta^a
   +\pi_a\partial^{\mu}A_{\mu}^a\right]
\end{equation}
where $\bar{\theta}_a$, $\theta^a$ are the Faddeev -- Popov ghosts, $D_{\mu}$ is a covariant derivative. The action is known to be BRST invariant. A direct demonstration of this fact can be found in any modern textbook on quantum field theory. The action (\ref{YMF}) includes second derivatives, and to construct the BRST charge one should used the Noether theorem generalized for theories with high order derivatives. In our case we have
\begin{equation}
\label{Noet.BRST}
\Omega_{Noether}=\int d^3x\left[\frac{\partial L}{\partial(\partial_0\phi^a)}\delta\phi^a
   +\frac{\partial L}{\partial(\partial _0\partial_{\mu}\phi^a)}\delta(\partial_{\mu}\phi^a)
   -\partial_{\mu}\left(\frac{\partial L}{\partial(\partial_0\partial_{\mu}\phi^a)}\right)\delta\phi^a\right],
\end{equation}
$\phi^a$ stands for field variables and ghosts. It gives the expression
\begin{equation}
\label{YM.BRST}
\Omega_{YM}=\int d^3x\left(-\theta^aD_ip_a^i-i\pi_a{\cal P}^a
   +\frac12\bar{{\cal P}}_ag f^a_{bc}\theta^b\theta^c\right)
\end{equation}
which coincides exactly with that obtained by the BFV prescription (\ref{BFV BRST}) after replacing the BFV ghosts by the Faddeev -- Popov ghosts; $p_a^i$, ${\cal P}^a$, $\bar{\cal P}_a$ are momenta conjugate to $A^a_i$, $\bar\theta_a$, $\theta^a$. But the situation in the gravitational theory is different.

\section{The BRST charge in the case of gravity}
In the case of gravity we deal with space-time symmetry, and we should take into account explicit dependence of the Lagrangian and the measure on space-time coordinates. The expression (\ref{Noet.BRST}) should be modified as
\begin{equation}
\label{grav.BRST}
\Omega_{grav}=\int d^3x\left[\frac{\partial L}{\partial(\partial_0\phi^a)}\delta\phi^a
   +\frac{\partial L}{\partial(\partial _0\partial_{\mu}\phi^a)}\delta(\partial_{\mu}\phi^a)
   -\partial_{\mu}\left(\frac{\partial L}{\partial(\partial_0\partial_{\mu}\phi^a)}\right)\delta\phi^a
   +\partial _0\left(L x^0\right)\right].
\end{equation}

We shall start from the simplest isotropic model with the action \cite{Shest1}:
\begin{equation}
\label{isotr}
S_{isotr}=\int dt\left[-\frac12\frac{a\dot a^2}N+\frac12 Na
 +\lambda\left(\dot N-\frac{df}{da}\;\dot a\right)
 +\bar\theta\frac d{dt}\left(-\dot N\theta-N\dot\theta+\frac{df}{da}\;\dot a\theta\right)\right].
\end{equation}
$N$ is the lapse function, $a$ is the scale factor. One can check that the action (\ref{isotr}) is not invariant under BRST transformations. However, the BRST invariance can be restored by adding to the action (\ref{isotr}) the additional term
\begin{equation}
\label{isotr-add}
S_1=\int dt\frac d{dt}\left[\bar\theta\left(\dot N-\frac{df}{da}\dot a\right)\theta\right].
\end{equation}
It contains only a full derivative and does not affect motion equations. We do not need any additional conditions to ensure the BRST invariance, for example, asymptotic boundary conditions for ghosts. The BRST charge constructed according to the Noether theorem (\ref{grav.BRST}) for the isotropic model would be
\begin{equation}
\label{isotr.BRST}
\Omega_{isotr}=-H\theta-\pi{\cal P},
\end{equation}
where
\begin{equation}
\label{isotr.Ham}
H=-\frac12\frac Na\left[p^2+2p\pi\frac{df}{da}+\pi^2\left(\frac{df}{da}\right)^2\right]
   -\frac12 Na+\frac1N\bar{\cal P}{\cal P},
\end{equation}
$p$ is the momentum conjugate to $a$, $\pi=\lambda+\dot{\bar\theta}\theta$ is the momentum conjugate to $N$, while $\bar{\cal P}$, $\cal P$ are ghost momenta. In the approach to Hamiltonian dynamics proposed in \cite{SSV1,SSV2} $H$ is the Hamiltonian in extended phase space. Thanks to the differential form of gauge condition in (\ref{isotr}), the Hamiltonian (\ref{isotr.Ham}) can be obtained by the usual rule
$H=\pi\dot N+p\dot a+\bar{\cal P}\dot\theta+\dot{\bar\theta}{\cal P}-L$ which is applicable to unconstrained systems. It is an important feature of this approach. Another its feature is that Hamiltonian equations in extended phase space are fully equivalent to Lagrangian equations, constraints and gauge conditions being true Hamiltonian equations. Making use of this, one can show that the charge (\ref{isotr.BRST}) generates correct transformations for all degrees of freedom, including gauge ones. By {\it correct} transformations I mean the ones that follow from transformations of metric tensor components
\begin{equation}
\label{g_transf}
\delta g_{\mu\nu}
 =\eta^{\lambda}\partial_{\lambda}g_{\mu\nu}
 +g_{\mu\lambda}\partial_{\nu}\eta^{\lambda}
 +g_{\nu\lambda}\partial_{\mu}\eta^{\lambda}
\end{equation}
taking into account a chosen parametrization of gravitational variables. For example,
\begin{equation}
\label{delta_N}
\delta N=\{N,\,\Omega_{isotr}\}=-\frac{\partial H}{\partial\pi}\theta-{\cal P}=-\dot N\theta-N\dot\theta,
\end{equation}
where we used the equation $\dot N=\displaystyle\frac{\partial H}{\partial\pi}$ (that is actually a differential form of the gauge condition $N=f(a)$), and the definition of the momentum ${\cal P}$ conjugate to $\bar\theta$.

The BRST charge constructed according to the BFV prescription (\ref{BFV BRST}) reads
\begin{equation}
\label{isotr.BFV.BRST}
\Omega^{BFV}_{isotr}=-T\theta-\pi{\cal P},
\end{equation}
where $T$ is the Hamiltonian constraint,
\begin{equation}
\label{Ham.constr.}
T=-\frac1{2a}p^2-\frac12 Na.
\end{equation}
The condition for physical states (\ref{quant BRST}) leads to the Wheeler -- DeWitt equation
\begin{equation}
\label{WDW}
\hat T|\Psi\rangle=0.
\end{equation}
The BFV charge (\ref{isotr.BFV.BRST}) fails to produce a correct transformation for the gauge variable $N$. At the same time, the condition (\ref{quant BRST}) with the Noether charge (\ref{isotr.BRST}), under the requirement of hermicity of Hamiltonian operator, does not lead to the Wheeler -- DeWitt equation.

We face the contradiction: on the one hand, at the classical level we have a mathematically consistent formulation of Hamiltonian dynamics in extended phase space which is equivalent to the Lagrangian formulation of the original theory, and the BRST generator constructed in accordance with the Noether theorem, that produces correct transformation for all degrees of freedom. On the other hand, at the quantum level our approach appears to be not equivalent to the BFV approach as well as the Dirac quantization scheme.

The investigation of more complicated models has confirmed the said above. Let us consider the generalized spherically-symmetric gravitational model \cite{Shest2} with the metric
\begin{eqnarray}
ds^2&=&\left[-N^2(t,r)+(N^r(t,r))^2V^2(t,r)\right]dt^2+2N^r(t,r)V^2(t,r)dt dr\nonumber\\
\label{mod.int}
&+&V^2(t,r)dr^2+W^2(t,r)\left(d\theta^2+\sin^2\theta d\varphi^2\right).
\end{eqnarray}
where $N^r=N^1$ is the only component of the shift vector. The model has two constraints and imitates the full theory of gravity much better. One can check that the sum of gauge-fixing and ghost parts of the action
\begin{eqnarray}
\label{gfix-act1}
S_{gauge}
&=&\int dt\int\limits_0^{\infty}dr\left[\lambda_0\left(\dot N-\frac{\partial f}{\partial V}\dot V
   -\frac{\partial f}{\partial W}\dot W\right)
  +\lambda_r\left(\dot N^r-\frac{\partial f^r}{\partial V}\dot V
   -\frac{\partial f^r}{\partial W}\dot W\right)\right];\\
S_{ghost}&=&\int dt\int\limits_0^{\infty}dr
   \left[\bar\theta_0\frac{d}{dt}\left(-\dot N\theta^0-N'\theta^r-N\dot\theta^0
    +N N^r(\theta^0)'\right.\right.\nonumber\\
&-&\left.\frac{\partial f}{\partial V}\left[-\dot V\theta^0-V'\theta^r-V(\theta^r)'-V N^r(\theta^0)'\right]
    -\frac{\partial f}{\partial W}\left[-\dot W\theta^0-W'\theta^r\right]\right)\nonumber\\
&+&\bar\theta_r\frac{d}{dt}\left(-\dot N^r\theta^0-(N^r)'\theta^r-N^r\dot\theta^0-\dot\theta^r
    +N^r(\theta^r)'+\frac{N^2}{V^2}(\theta^0)'+(N^r)^2(\theta^0)'\right.\nonumber\\
\label{ghost-act1}
&-&\left.\left.\frac{\partial f^r}{\partial V}\left[-\dot V\theta^0-V'\theta^r-V(\theta^r)'-V N^r(\theta^0)'\right]
    -\frac{\partial f^r}{\partial W}\left[-\dot W\theta^0-W'\theta^r\right]\right)\right]
\end{eqnarray}
is not invariant under BRST transformations. To ensure its BRST invariance we have to add to the action the following terms (compare with (\ref{isotr-add})):
\begin{eqnarray}
S_2&=&\int dt\int\limits_0^{\infty}dr
   \left(\frac d{dt}\left[\bar\theta_0\left(\dot N-\frac{\partial f}{\partial V}\dot V
     -\frac{\partial f}{\partial W}\dot W\right)\theta^0\right]
    +\frac d{dr}\left[\bar\theta_0\left(\dot N-\frac{\partial f}{\partial V}\dot V
     -\frac{\partial f}{\partial W}\dot W\right)\theta^r\right]\right.\nonumber\\
\label{spher-add}
&+&\left.\frac d{dt}\left[\bar\theta_r\left(\dot N^r-\frac{\partial f^r}{\partial V}\dot V
     -\frac{\partial f^r}{\partial W}\dot W\right)\theta^0\right]
    +\frac d{dr}\left[\bar\theta_r\left(\dot N^r-\frac{\partial f^r}{\partial V}\dot V
     -\frac{\partial f^r}{\partial W}\dot W\right)\theta^r\right]\right).
\end{eqnarray}

The BRST charge constructed according to the Noether theorem (\ref{grav.BRST}) for the spherically-symmetric model is
\begin{eqnarray}
\Omega_{spher}&=&\int\!dr\left[-{\cal H}\theta^0-P_V V'\theta^r
   -P_N\frac{\partial f}{\partial V}V'\theta^r
   -P_{N^r}\frac{\partial f^r}{\partial V}V'\theta^r-P_W W'\theta^r\right.\nonumber\\
&-&P_N\frac{\partial f}{\partial W}W'\theta^r
   -P_{N^r}\frac{\partial f^r}{\partial W}W'\theta^r
   -P_V V N^r(\theta^0)'-P_N\frac{\partial f}{\partial V} V N^r(\theta^0)'\nonumber\\
&-&P_{N^r}\frac{\partial f^r}{\partial V} V N^r(\theta^0)'
   -P_V V(\theta^r)'-P_N\frac{\partial f}{\partial V} V(\theta^r)'
   -P_{N^r}\frac{\partial f^r}{\partial V} V(\theta^r)'\nonumber\\
\label{spher.BRST}
&-&\left.\bar P_{\theta^0}(\theta^0)'\theta^r
   -\bar P_{\theta^r}(\theta^r)'\theta^r
   -P_N P_{\bar\theta_0}-P_{N^r}P_{\bar\theta_r}
   -\frac{N W W'(\theta^0)'}V\right],
\end{eqnarray}
$\cal H$ is a Hamiltonian density in extended phase space, its explicit form is given in \cite{Shest2}. It has been also demonstrated in \cite{Shest2} based on the equivalence of the Lagrangian and Hamiltonian dynamics for this model that the BRST charge (\ref{spher.BRST}) generates {\it correct} transformations (in the sense explained above) for physical, gauge and ghost degrees of freedom. Nevertheless, its structure differs from that of the BFV charge.

Nothing prevents us from constructing Hamiltonian dynamics in extended phase space and the BRST charge for the full gravitational theory following the method outlined above. One can use a gauge condition in a general form,
$f^{\mu }(g_{\nu\lambda})=0$. Its differential form introduces the missing velocities and {\it actually extends} phase space, so that the gauge fixing and ghost parts of the action will be
\begin{eqnarray}
\label{gfix-act2}
S_{(gauge)}&=&\int d^4 x\,\lambda_{\mu}\frac{d}{dt}f^{\mu}(g_{\nu\lambda})
   =\int d^4 x\,\lambda_{\mu}\left(\frac{\partial f^{\mu}}{\partial g_{00}}\dot g_{00}
     +2\frac{\partial f^{\mu}}{\partial g_{0i}}\dot g_{0i}
     +\frac{\partial f^{\mu}}{\partial g_{ij}}\dot g_{ij}\right);\\
\label{ghost-act2}
S_{(ghost)}&=&-\int d^4 x\,\bar\theta_{\mu}\frac{d}{dt}
  \left[\frac{\partial f^{\mu}}{\partial g_{\nu\lambda}}
   \left(\partial_{\rho}g_{\nu\lambda}\theta^{\rho}
     +g_{\lambda\rho}\partial_{\nu}\theta^{\rho}
     +g_{\nu\rho}\partial_{\lambda}\theta^{\rho}\right)\right].
\end{eqnarray}
It is not difficult to check that the additional term ensuring BRST invariance of the action in this general case reads (compare with (\ref{isotr-add}), (\ref{spher-add})):
\begin{equation}
\label{gen-add}
S_3=\int d^4 x\,\partial_{\mu}\left[\bar{\theta}_{\nu}
    \frac d{dt}f^{\nu}(g_{\lambda\rho})\theta^{\mu}\right].
\end{equation}

The calculation of the BRST charge for the full gravitational theory is rather tedious and has not been finished yet. However, relying upon the two models discussed above, we can expect that the structure of the BRST charge may also be different from the one predicted by Batalin, Fradkin and Vilkovisky.

\section{Discussion}
Therefore, one should inquire about a physical meaning of the selection rules (\ref{quant BRST}) (in the BFV approach) or (\ref{quant Dirac}) (in the Dirac approach) as well as asymptotic boundary conditions. In quantum field theory with asymptotic states their meaning is quite clear: in asymptotic states interactions are negligible, and these states must not depend on gauge and ghost variables which are considered as non-physical. But ghost fields cannot be excluded in an interaction region. In the gravitational theory, except some few situations, we need to explore states inside the interaction region. The simplest example of a system without asymptotic states is a closed universe, not to mention a universe with more complicated topology. Also, we would like to reach a better understanding of quantum processes in the neighborhood of a black hole. Then, what would be a definition of physical states in such cases? To my mind, today we have no satisfactory answer for this question, though mathematics provides reasonable grounds to put it. The definition of physical states seems to be very important for our searching for a true way to quantize Gravity.

\end{document}